\documentclass[onecolumn,superscriptaddress,nobalancelastpage,12pt,pra,a4paper]{revtex4}
\usepackage{graphicx}
\usepackage{dcolumn}
\usepackage{bm}
\usepackage{verbatim}
\usepackage{soul}
\usepackage[dvipsnames]{xcolor}
\colorlet{RED}{red}
\usepackage{braket}
\usepackage{dsfont}
\usepackage{amsmath,subfigure,float}
\usepackage{url}
\usepackage{ctable}
\usepackage{soul}
\usepackage[USenglish]{babel}
\usepackage{titlesec}
\addto{\captionsUSenglish}{}

\newcommand{\E}{\mathrm{e}}

\newcommand{\rhobf}{\boldsymbol{\mathbf{\rho}}}
\newcommand{\rhoc}{\boldsymbol{\mathbf{\rho}}_{\mathrm{c}}}
\newcommand{\rhoo}{\boldsymbol{\mathbf{\rho}}_{\mathrm{o}}}
\newcommand{\rmc}{\mathrm{c}}
\newcommand{\rmd}{\mathrm{d}}
\newcommand{\rmo}{\mathrm{o}}
\newcommand{\rmu}{\mathrm{u}}
\newcommand{\rmG}{\mathrm{G}}
\newcommand{\rmV}{\mathrm{V}}

\newcommand{\wP}{w_\mathrm{p}}
\newcommand{\qd}{\mathbf{q}_\mathrm{d}}
\newcommand{\qu}{\mathbf{q}_\mathrm{u}}
\newcommand{\lamd}{\lambda_\mathrm{d}}
\newcommand{\lamu}{\lambda_\mathrm{u}}
\newcommand{\ESF}{\mathrm{ESF}}
\newcommand{\PSF}{\mathrm{PSF}}

\newcommand{\mm}{\;\mathrm{mm}}
\newcommand{\um}{\;\mathrm{\mu m}}
\newcommand{\nm}{\;\mathrm{nm}}

\DeclareMathOperator{\erf}{erf}
\DeclareMathOperator{\sinc}{sinc}

\usepackage{ulem} 

\titleformat{\section}{\normalfont\large\bfseries}{\thesection}{10pt}{}
\titleformat{\subsection}{\normalfont}{\thesubsection}{10pt}{\ul}
\titlespacing{\subsection}{0pt}{\parskip}{4pt}
\titleformat{\subsubsection}{\normalfont\large}{\thesubsection}{10pt}{}
\titlespacing{\subsubsection}{0pt}{\parskip}{4pt}

\usepackage{xr}
\usepackage{hyperref}

\makeatletter
\newcommand*{\addFileDependency}[1]{
  \typeout{(#1)}
  \@addtofilelist{#1}
  \IfFileExists{#1}{}{\typeout{No file #1.}}
}
\makeatother

\newcommand*{\myexternaldocument}[1]{%
    \externaldocument{#1}%
    \addFileDependency{#1.tex}%
    \addFileDependency{#1.aux}%
}

\myexternaldocument{SM}

\begin{document}
\title{Experimental analysis on image resolution of quantum imaging with undetected light through position correlations}
\author{Marta Gilaberte Basset$^{1,2,\dagger,*}$, René Sondenheimer$^{1,3,\dagger,**}$, Jorge Fuenzalida$^{4}$, Andres Vega$^{2}$, Sebastian T{\"o}pfer$^{4}$, Elkin A. Santos$^{2}$, Sina Saravi$^{2}$, Frank Setzpfandt$^{1,2}$, Fabian Steinlechner$^{1,2}$, and Markus Gr{\"a}fe$^{1,2,4}$
\vspace{1em}
\\
\it $^{1}$Fraunhofer Institute for Applied Optics and Precision Engineering IOF,
\it Albert-Einstein-Str. 7, 07745, Jena, Germany. \\ \vspace{0.5em}
\it $^{2}$Friedrich-Schiller-University Jena, Institute of Applied Physics, Abbe Center of Photonics,\\ 
\it Albert-Einstein-Str. 6, 07745, Jena, Germany.\\ \vspace{0.5em}
\it $^{3}$Friedrich-Schiller-University Jena, Institute of Condensed Matter Theory and Optics,\\ 
\it Max-Wien-Platz 1, 07743 Jena, Germany.\\ 
\it $^{4}$Institute of Applied Physics, Technical University of Darmstadt, Schloßgartenstraße 7, 64289 Darmstadt, Germany\\ 
\vspace{0.5em}
$^{\dagger}$Both authors contributed equally.\\
$^{*}$marta.gilaberte.basset@iof.fraunhofer.de\\
$^{**}$rene.sondenheimer@iof.fraunhofer.de\\
}

\date{\today}

\maketitle

\newpage
\section*{Abstract}

Image resolution of quantum imaging with undetected photons is governed by the spatial correlations existing between the photons of a photon pair that has been generated in a nonlinear process. These correlations allow for obtaining an image of an object with light that never interacted with that object. Depending on the imaging configuration, either position or momentum correlations are exploited. We hereby experimentally analyse how the crystal length and pump waist affect the image resolution when using position correlations of photons that have been generated via spontaneous parametric down conversion in a nonlinear interferometer. Our results support existing theoretical models for the dependency of the resolution on the crystal length. In addition, we probe the resolution of our quantum imaging scheme for varying pump waists over one order of magnitude. This analysis reveals the intricate dependency of the resolution on the strength of the correlations within the biphoton states for parameter combinations in which the crystal lengths are much larger than the involved photon wavelengths. We extend the existing models in this parameter regime to properly take nontrivial effects of finite pump waists into account and demonstrate that they match the experimental results.





\newpage

\section{Introduction}
In recent years, quantum imaging techniques have proven to be a very useful tool to overcome classical limitations~\cite{GilaberteBasset2019_Rev, Moreau2019}. For instance, when imaging at wavelengths outside the visible range, detection technologies are limited especially for low-light level applications, such as occurring in life sciences~\cite{Kviatkovsky2021_Microscope}. Quantum imaging with undetected light (QIUL)~\cite{Lemos2014} is a technique that overcomes these detection limitations exploiting the capabilities of nonlinear interferometers \cite{Yurke1986, Herzog1995, Chekhova2016}. It is based on the quantum interference effect of induced coherence~\cite{Zou1991,Wang1991}, and exploits the spatial correlations existing between two photons for example generated via spontaneous parametric down-conversion (SPDC), to create an image of an object with light that did not illuminate it. 
This nonlinear process can be engineered to generate one beam at the desired probe wavelength for the sample, and the other beam, containing correlated partner photons, at the visible range to ease the detection. Therefore, the interest in understanding quantum imaging systems has rapidly grown not only for imaging applications \cite{Lahiri2015_TheoryLemos, Lahiri2017, Kviatkovsky2021_Microscope}, but also for holography~\cite{Topfer2022, Haase_2023}, spectroscopy~\cite{Kalashnikov2016, Paterova2017_spectroscopy, Lindner2022}, and optical coherence tomography~\cite{Valles2018, Paterova2018_OCT}. 

Image resolution is one of the main parameters that describes the quality of an imaging system, which for QIUL is governed by the spatial correlations of the photons. 
Several works have experimentally exploited the momentum anti-correlations of SPDC biphoton states, i.e. imaging at the far-field plane (Fourier plane) of the nonlinear crystal~\cite{Lemos2014, Paterova2020_Semiconductor, Paterova2020_Microscopy, Paterova2021,Kviatkovsky2021_Microscope, GilaberteBasset2021_Video-rate, Topfer2022, Fuenzalida2023_Distillation}. 
Alternatively, one can also obtain the image of an object that is placed at the near-field plane (image plane) of the nonlinear crystal. When this is the case, the imaging system exploits position correlations of the photons~\cite{Viswanathan2021_July}.
Recently, QIUL has been implemented with the near-field configuration for the first time and, thus, demonstrating its experimental viability~\cite{Kviatkovsky2022_Position}.
Exploiting position correlations is of particular interest due to the fact that the degree of correlation between the two photons of an SPDC pair does not depend on the pump beam spatial coherence~\cite{Defienne2019, Zhang2019}. That relaxes the requirements on the pump source, providing more flexibility to engineer a quantum imaging system.

The role of the parameters of the two-photon source on image resolution (pump waist, crystal length, and wavelengths of the down-converted photons) has been analysed for both cases momentum~\cite{Fuenzalida2022} and position~\cite{Viswanathan2021_November, Vega2022} correlations.

These works derived resolution limits for specific parameter regimes within different approximations that were specifically designed for the precise parameter regime under consideration. For instance, the crystal length can be neglected within a thin-crystal approximation~\cite{Monken1998} in the far-field configuration~\cite{Fuenzalida2022}. 
By contrast, the impact of a finite pump waist can usually be neglected for near-field configurations but the crystal length plays the dominant role for image resolution. In particular, it has been shown that shorter crystal lengths improve the resolution within the paraxial regime~\cite{Viswanathan2021_November}. However, this improvement reaches a lower bound given by the diffraction limit. At this limit, the resolution is governed by the longer wavelength of the photon pair. For a detailed analysis providing a general model for such effects also beyond the commonly used paraxial regime, we refer to~\cite{Vega2022}. Additionally, sub-diffraction resolution imaging might be achieved by exploiting evanescent modes existing within wavelength-range distances~\cite{Santos2022}.

While the theory predictions in the far field have been experimentally demonstrated~\cite{Fuenzalida2022}, this task remains missing in the near field. In this work, we experimentally study the resolution of QIUL for different parameter regimes based on position correlations for the first time. We demonstrate that in this configuration the main parameter governing the spatial resolution for sufficiently large crystals is the crystal length as in agreement with the theory \cite{Viswanathan2021_November, Vega2022}.
Furthermore, we also vary the pump waist to show that it does not influence the resolution over a broad parameter range. In particular, if the wavelengths are in the visible or near infrared regime and the crystal length is of the order of millimeters, the resolution stays almost unaffected for pump waists $\gtrsim100\um$.
However, we observe slight deviations for strongly focused pump beams. 

To properly account for these effects, the existing theoretical model for the system needs to be extended. Although recently developed numerical techniques~\cite{Vega2022} could also account for such effects, we generalize the existing analytical model for resolution limits in the near field \cite{Viswanathan2021_November} to derive an analytical dependency of the image resolution on the pump waist as well as the crystal length. 
This investigation also allows us to directly connect the image resolution with the strength of the quantum correlations encoded in the biphoton states. Moreover, it reveals that different physical information is stored in the visibility and the image function that might be used to describe the imaging system. Resolution can be determined via characteristic spreads quantifying the blurring seen in an image of an object. Spreads extracted from both functions almost coincide for sufficiently large pump waists in the near-field configuration such that the resolution limit can be obtained either from amplitude images (image function) or from visibility images (visibility). However, they deviate for decreasing pump waists showing that the correlation information between the photons is only properly reflected in the visibility. We will show that these effects are corroborated in the obtained experimental results. As a side product, our analysis provides a new quantity to assess the quality of the imaging setup without needing any information about involved magnifications. In case the experimentally measured data stays sufficiently close to the corresponding theory predictions, we are able to introduce a tool to extract an estimator for the magnification value of the imaging configuration without the need to directly measure it.

\section{Experimental setup}
\label{sec:experimental setup}
\begin{figure}[t]
	\centering
		\includegraphics[width=0.76\linewidth]{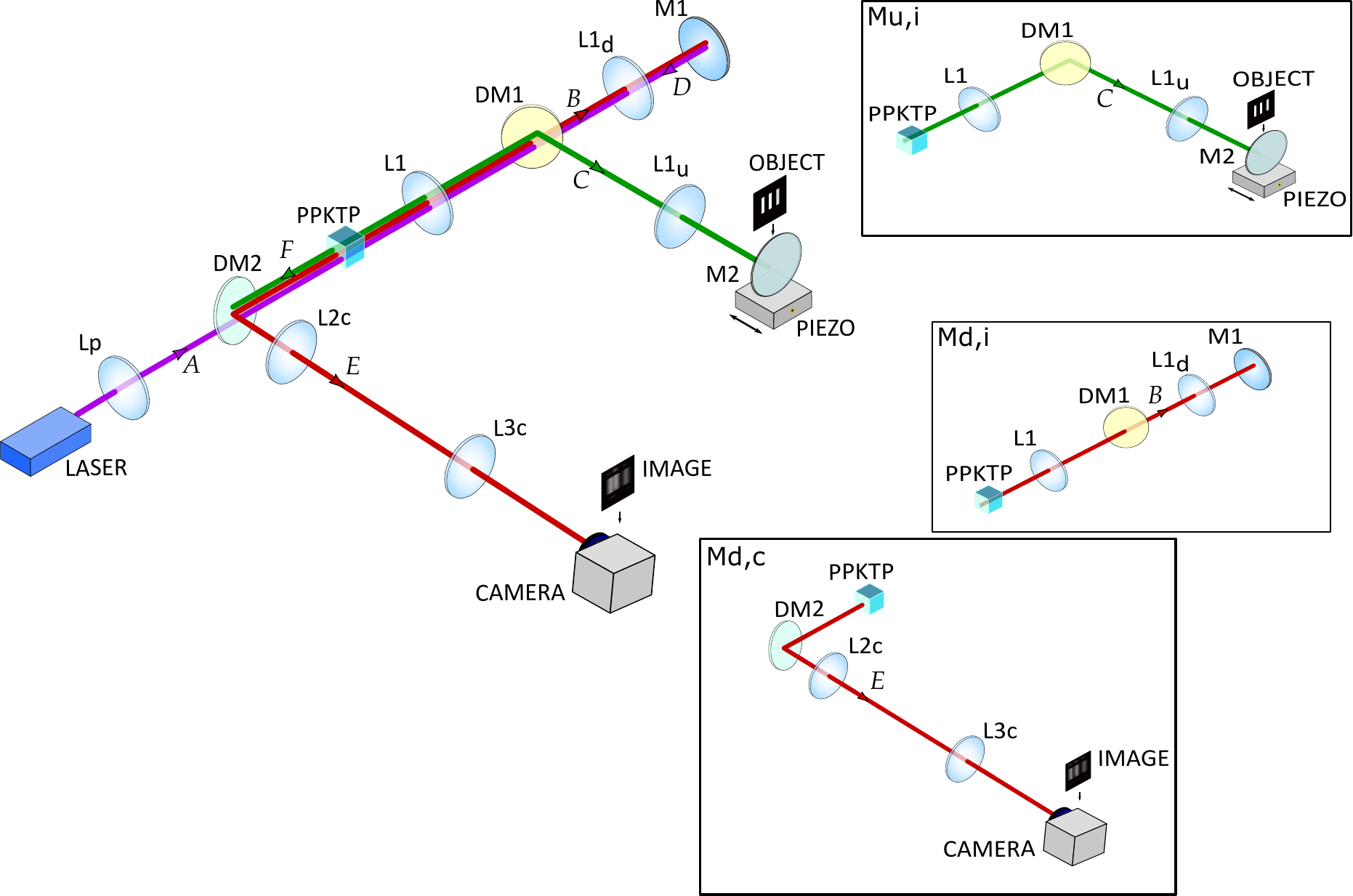}    \caption{\textbf{Near-field configuration setup.} A pair of down-converted photons is generated at the ppKTP crystal in either the forward pass of the pump (path A) through the crystal or in the backward pass (path D). The pump beam is focused at the crystal with $\mathrm{L}_{\mathrm{p}}$. We change $\mathrm{L}_{\mathrm{p}}$ focal length and its distance to the crystal accordingly, to generate different pump waists at the center of the crystal. The undetected beam is reflected with DM1 towards the object while the detected and the pump beam are transmitted together. The DM2 reflects the detected beam towards the camera. The interferometer phase is varied by changing the position of the mirror M2 with a piezo translation stage. The lenses $\mathrm{L}1$ and $\mathrm{L}1_\rmu$ (with equal focal lengths, $f_1=125\mm$) image the generated modes in the crystal onto the object plane, and back onto the crystal after being reflected back by the mirror M2. This configuration allows to exploit the position correlations of the photon pairs. Lens $\mathrm{L}1_\rmd$ has also focal length $f_1$. The lenses $\mathrm{L}2_\mathrm c$ ($f_2=75\mm$) and $\mathrm{L}3_\mathrm c$ ($f_3=200\mm$) image, and magnify, the object at the camera. Long pass and interference filters guarantee that only the $910\nm$ wavelength is detected. The insets show the different magnification systems present: magnifications of the undetected (detected) beam inside the interferometer ($M_\mathrm{u(d),i}$), and the magnification of the detected beam before being detected on the camera ($M_\mathrm{d,c}$).}
	\label{fig:1_NFsetup}
\end{figure}

The experimental setup (Fig.~\ref{fig:1_NFsetup}) consists of an SU(1,1) nonlinear interferometer where a 4f system of lenses ensures that the object lies in the image plane (near field) of the crystal. This plane is then imaged into the camera through a different 4f lens system. In this way, position correlations enable the formation of the image ~\cite{Viswanathan2021_July}. For more details on the systems of lenses and the imaging configuration used, see Fig. \ref{fig:1_NFsetup_simple} in App.~\ref{app:imaging_configuration}.

A pump beam of $96\;\mathrm{mW}$ pump power and pump wavelength $\lambda_{\mathrm{p}} = 405\nm$ is focused with lens $\mathrm{L}_{\mathrm{p}}$ into a type-0 ppKTP crystal that generates a pair of correlated photons through SPDC at $730\nm$ and $910\nm$ wavelengths either during the forward propagation of a pump photon through the crystal (path $\mathrm A$) or when it passes through the crystal in the backward direction (path $\mathrm D$) after being reflected back by mirror M1. 
We refer to the light with wavelength $\lamu = 910\nm$ as undetected ($\mathrm u$) because it is never detected, although being the one illuminating the object. By contrast, the photons with wavelength $\lamd = 730\nm$ are directed towards the camera but never interact with the object. Therefore we denote this beam as the detected beam ($\mathrm d$). The camera used for detection is a Prime BSI Scientific CMOS from Teledyne Photometrics with a pixel size of $6.5\um$. Because of the sufficiently low pump power, the down-conversion process occurs in the low gain regime and we can consider only one pair of down-converted photons (either forward or backward generated) to be present at a time in the interferometer. The probability amplitudes of the SPDC emission generated in the first and second passage through the nonlinear crystal are superposed and exhibit interference when indistinguishable. The required indistinguishability is achieved by careful alignment of the forward and backward beams which erases the which-path information. The interference pattern observed from the detected photons contains information of an object in the undetected beam path ($\mathrm C$) due to the induced coherence without induced emission effect \cite{Wang1991, Zou1991}.

Using this quantum phenomenon, the image formation for QIUL works as follows. 
An undetected photon in path $\mathrm C$ (Fig.~\ref{fig:1_NFsetup}) with transverse wave vector $\qu$ and transverse position $\rhobf_\mathrm{u}$ interacts with an object placed at the image plane of the nonlinear crystal at the transverse position $\rhoo = M_\rmu \rhobf_\mathrm{u}$ where $M_\rmu$ is the total magnification obtained by photons of the undetected arm. This spatial information is linked to a photon with transverse position ($\rhobf_\mathrm{d}$) in the detected beam due to the correlations of the SPDC biphoton states originating from the common creation event of the photon pair. This photon is detected at the camera position $\rhoc = M_\rmd \rhobf_\mathrm{d}$ with $M_\rmd$ denoting the total magnification for light in the detected path~\cite{Herzog1995}. 
Therefore, a position on the object $\rhoo$ is directly related to a position on the camera $\rhoc$.  

The optimal visibility of the interference generated in such a scheme is achieved by accurate alignment of the optical components for indistinguishability of the beams (and to fulfill the imaging conditions), as well as precisely matching the interferometric arms to the same optical length. Image resolution is affected by the precision of this alignment as well.

The mirror M2 is mounted on top of a piezo stage to allow for the scanning of different interferometric phases, which allows us to apply the digital phase-shifting holography (DPSH) technique to extract images with amplitude and phase information of the object~\cite{Topfer2022}. Amplitude images obtained from DPSH can be directly related to the value of the image function $G(\rhoc)$ at each camera pixel.
The image function has been introduced as the difference of the maximum ($I_{\mathrm{max}}$) and minimum intensity ($I_{\mathrm{min}}$) at each position in the camera plane~\cite{Viswanathan2021_November},
\begin{align}
    G(\rhoc) = I_{\mathrm{max}}(\rhoc) - I_{\mathrm{min}}(\rhoc).
\label{eq:imageFunction-Def}
\end{align}

Visibility, given by
\begin{align}
    V(\rhoc) = \frac{I_{\mathrm{max}}(\rhoc) - I_{\mathrm{min}}(\rhoc)}{I_{\mathrm{max}}(\rhoc) + I_{\mathrm{min}}(\rhoc)},
\label{eq:visibility-Def}
\end{align}
at each pixel is also extracted as an image, which we call visibility image. The latter can be used to analyse the system resolution and the strength of the position correlations. 

The lenses inside the interferometer ($\mathrm{L}1$, $\mathrm{L}1_\mathrm d$ and $\mathrm{L}1_\mathrm u$) introduce no magnification ($M_\mathrm{u,i}=M_\mathrm{d,i}=1$) when the system of lenses is perfectly positioned. 
$M_\mathrm{u(d),i}$ is the magnification of the lens system in the interferometer undetected (detected) beam path. The detected beam passes through a second magnification system on its way to the camera ($M_\mathrm{d,c}$) consisting of lenses $\mathrm{L}2_\mathrm c$ and $\mathrm{L}3_\mathrm c$ introducing a magnification of 2.67. Magnifying the image allows us to have more precision in the measurements, due to the pixel size. The total magnification seen by the detected beam is then $M_\mathrm d = M_\mathrm{d,i} M_\mathrm{d,c}$. In practice, ensuring these precise magnification value ($M_\rmd = 2.67$) is a challenging task and often difficult to realize. In the following, we will elaborate on how we account for the impact of not ideally positioned lenses by extracting the relevant magnification value, here $M_\rmd$, for each tested configuration from a nonlinear fit of the intensity and visibility profiles generated by a sharp edge on the camera. This newly introduced routine allows for higher accuracy than our experimental evaluation of the magnification, see App~\ref{app:MagMeas} for more details. 

The actual resolution of the implemented quantum imaging scheme with undetected photons depends on various quantities coming either from the quantum nature of the underlying SPDC process, i.e. the correlation strength of the biphoton state, or the classical imaging system in terms of image formation and magnification. In order to isolate the impact of the underlying quantum correlations depending on the crystal length $L$ and pump waist $\wP$, we need to know the precise total magnification from the lens system. This is important because the spreads measured in the camera plane explicitly depend on $M_\rmd$ in a multiplicative fashion as the detected photons creating the image precisely go through the corresponding lens system. Therefore, we can construct magnification-adjusted spreads $\Delta_\rmV = \frac{\Delta_{\rmV,\rmc}}{M_\rmd}$ and $\Delta_\rmG = \frac{\Delta_{\rmG,\rmc}}{M_\rmd}$ where $\Delta_{\rmV,\rmc}$ and $\Delta_{\rmG,\rmc}$ denote the spreads measured in the camera plane. The subscripts $\rmV$ and $\rmG$ refer to the fact whether the spreads where obtained from visibility or amplitude (image function) images, respectively. In order to extract information in the object plane, we multiply the magnification adjusted visibility spread by the magnification of the undetected arm, $\Delta_{\rmo} = M_\rmu\Delta_{\rmV}$, i.e. the total magnification of the system, relating camera and object planes, is given by $\frac{\Delta_{\rmV,\rmc}}{\Delta_{\rmo}} = \frac{M_\rmd}{M_\rmu}$. The magnification-adjusted spreads provide information about the imaging system that root purely in the quantum nature of the implemented scheme and factor out any impact induced by the classical part, e.g. optical aberrations, imaging system misalignments, or magnifications. This is equivalent to realize a system where the lens configuration does not imply any magnification at all. As the main focus of our work will be on the impact of the quantum correlations on the spatial resolution, we will focus on the magnification-adjusted spreads in the following.

Due to a low manufacturing precision of our target object for the measurement of the magnification, the experimental results obtained suffered from big error bars (see App.~\ref{app:MagMeas} for more details). In order to minimize the uncertainty coming from the magnification measurement, we propose a different strategy that allows us to construct an estimator for the magnification present in the system. First, we introduce a new parameter to quantify the quality of the experimental results without the need of knowing the system magnification. As the spreads, either obtained from image function or visibility, measured in the camera depend only linearly on $M_\rmd$, we study their ratio 
\begin{align}
    \frac{\Delta_{\rmG,\rmc}}{\Delta_{\rmV,\rmc}} = \frac{\Delta_{\rmG}}{\Delta_{\rmV}}
\label{eq:ratio-M-independent}
\end{align}
which is a magnification independent quantity by construction. Although this quantity cannot be related to the resolution of the system, we can use it to estimate the quality of the correlations and the overall alignment required to generate induced coherence. The advantage of this ratio is given by the fact that we are able to compare pure experimentally obtained data (left-hand side of Eq.~\eqref{eq:ratio-M-independent}) to values that can be predicted by theory (right-hand side of Eq.~\eqref{eq:ratio-M-independent}). In case the experimentally obtained ratio stays close to the theory prediction, we can use the following functional dependency as a fit for the experimentally obtained data for the image function (cf. Sec.~\ref{sec:theory} for a derivation)
\begin{align}
 G_{\ESF}(x_\mathrm{c}) &= \exp{ \left\{ -\frac{4\pi (\lamd + \lamu)}{\lamd^2 L + 2\pi\wP^2 (\lamd + \lamu)} \frac{x_\mathrm{c}^2}{M_\rmd^2} \right\} } \times \notag\\
 &\qquad\quad \times
 \left[ 1 - \erf\bigg\{ \frac{\sqrt{2} [\lambda_\rmd\lambda_\rmu L - 2\pi\wP(\lamd + \lamu)]}{\sqrt{[\lambda_\rmd^2 L + 2\pi\wP(\lamd + \lamu)]L}\wP(\lamd + \lamu)} \frac{x_\mathrm{c} - M_\rmu\tilde{x}_\rmo}{M_\rmd} \bigg\} \right]
 \label{eq:magnification-G}
\end{align}
and visibility 
\begin{align}
 V_{\ESF}(x_\mathrm{c}) &= \frac{1}{2}
 \left[ 1 - \erf\bigg\{ \frac{\sqrt{2} [\lambda_\rmd\lambda_\rmu L - 2\pi\wP(\lamd + \lamu)]}{\sqrt{[\lambda_\rmd^2 L + 2\pi\wP(\lamd + \lamu)]L}\wP(\lamd + \lamu)} \frac{x_\mathrm{c} - M_\rmu\tilde{x}_\rmo}{M_\rmd} \bigg\} \right]
 \label{eq:magnification-V}
\end{align}
to estimate the magnification of the detected photon beam $M_\rmd$. The subscript $\ESF$ denotes that we have evaluated the image function and visibility for a sharp edge model. Here, $x_\rmc$ denotes the horizontal coordinate in the camera plane and $\tilde{x}_\rmo$ accounts for a potential displacement of the object from the optimal position at the center of the undetected light beam. The impact of the parameter combination $M_\rmu\tilde{x}_\rmo$ on the resolution in the camera plane can also be analyzed by this fit routine, i.e. we are using a two parameter fit with fit parameters $M_\rmd$ and $M_\rmu\tilde{x}_\rmo$.

Strictly speaking, both fits (image function and visibility) should give the same magnification value. However using the fits, we implicitly assume that the underlying theoretical model matches perfectly the experimental realization. Due to experimental uncertainties, e.g. in the alignment or the determination of the other parameters of the system, there can be an ambivalence in the extraction of the magnification parameter $M_\rmd$. This can potentially result in a deviation of $M_\rmd$ extracted from Eq.~\eqref{eq:magnification-G} from $M_\rmd$ obtained from Eq.~\eqref{eq:magnification-V}. Nonetheless, as long as the magnification independent ratio $\Delta_{\rmG,\rmc}/\Delta_{\rmV,\rmc}$ stays close to the theory prediction $\Delta_{\rmG}/\Delta_{\rmV}$ which implies that the theoretical model fits sufficiently good the experimental realization, the magnification extracted from both functions will be (almost) the same. Therefore, we use the average of these two values as an estimator of $M_\rmd$ if this is the case. Note that this careful comparison is necessary. Using only one fit as an estimator of $M_\rmd$ could lead to wrong magnification values as a change in $M_\rmd$ can be compensated in a change of other system parameters. Then, one would extract an incorrect magnification value designed in such a way that the resulting resolution imitates the theory prediction.

\section{Experimental results}
\label{sec:results}
In this section, we present the results on image resolution when exploiting position correlations in QIUL. To evaluate the effect of the crystal thickness on the resolution, the measurements were performed with three crystals of the same characteristics but different lengths $L$ ($2\mm$, $5\mm$, and $10\mm$). Additionally, for each crystal, the resolution of the system is evaluated for different pump waists $\wP$ ($50\um$, $142\um$, $214\um$, and $308\um$).

\begin{figure}
	\centering
		\includegraphics[width=0.5\linewidth]{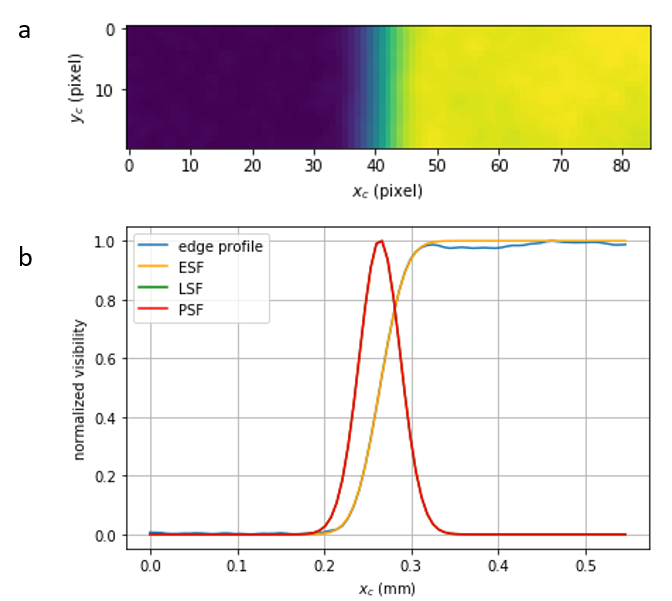}
	\caption{\textbf{Spread of an edge analysis.} a) Cut from the visibility image taken to analyse the spread of an edge blocking the left side of the beam. b) The edge profile (blue) is extracted from the image, and fitted with an ESF (orange). From the ESF, the LSF (green) and PSF (red) are calculated. LSF and PSF overlap. Example extracted from data taken with a $2\mm$ long crystal and $214\um$ pump waist.}
 	\label{fig:LSF_PSF}
\end{figure}

The resolution power of our system is obtained through the analysis of the edge response of the system to a sharp edge. The object is a blade of a knife edge placed at the image plane of the crystal (near-field configuration), right in front of the mirror on the signal arm (M2 in Fig.~\ref{fig:1_NFsetup}) such that it is imaged parallel to the vertical $y_{\rmc}$-axis in the camera plane (see Fig.~\ref{fig:LSF_PSF}). 

The edge response is evaluated from both, amplitude and visibility images at the camera plane. In order to do this, we first analyze the integrated intensities per pixel row to determine the $y_\rmc$ position with maximum intensity for each amplitude image. By doing this, we determine the optimal position where the detected beam has the strongest impact which minimizes errors induced by our theoretical approximations. Then, we fit the image and visibility functions evaluated for a sharp edge model, i.e. the corresponding edge spread functions (ESFs), to the experimentally obtained amplitude and visibility edge profile for the pixel row with maximum intensity, respectively. 
Many classical imaging schemes are linear and stationary/isoplanatic such that the impulse response function depends only on coordinate differences between the object and camera planes. In this case, the derivative of the ESF is equivalent to the line spread function (LSF) which, in turn, is directly related to the point spread function (PSF) when considering a Gaussian profile for the illumination \cite{Blackledge2005} (see Fig.~\ref{fig:LSF_PSF}).

Exploiting position correlations for QIUL, these relations are fulfilled for visibility within our approximation as well as for the image function (amplitude) to a good approximation if the pump waist is sufficiently large. However, we would like to emphasize at this point that for smaller pump waists, the derivative of the ESF will not coincide with the LSF for the image function as the system is no longer isoplanatic which can be directly inferred from the joint probability distribution of detected and undetected photons, see Sec.~\ref{sec:theory}. We also observe that for sufficiently large pump waists ($\gtrsim100$ $\mu$m in our configuration) the analysis of amplitude images to extract the system resolution power gives similar results as visibility images, but they strongly differ for smaller pump waists. Only in the particular parameter regime where the conditions $\wP^2 \gg \frac{\lamu^2 L}{\lamd + \lamu}$ and $\wP^2 \gg \frac{\lamd^2 L}{\lamd + \lamu}$ are fulfilled, the image function might be used to determine the image resolution to a good approximation. We will elaborate on these points in detail in Sec.~\ref{sec:theory}.


The resolution of an imaging system can be heuristically defined in various ways. Here, we follow the practical convention that we analyze the spread of the PSF at the point where its intensity decays to $1/\E$ in order to directly compare our results with previous works~\cite{Fuenzalida2022,Viswanathan2021_November}. 
This definition can in most cases also be transferred to a $1/\E$-width of the LSF or an $24/76$-knife-edge width of the ESF being defined as the distance between the points of the measured curve that are 24\% and 76\% of the maximum value.
While this analogy holds for visibility images, it is not the case for amplitude images (also see Sec.~\ref{sec:theory} for a detailed discussion). 

\begin{figure}
	\centering
		\includegraphics[width=0.8\linewidth]{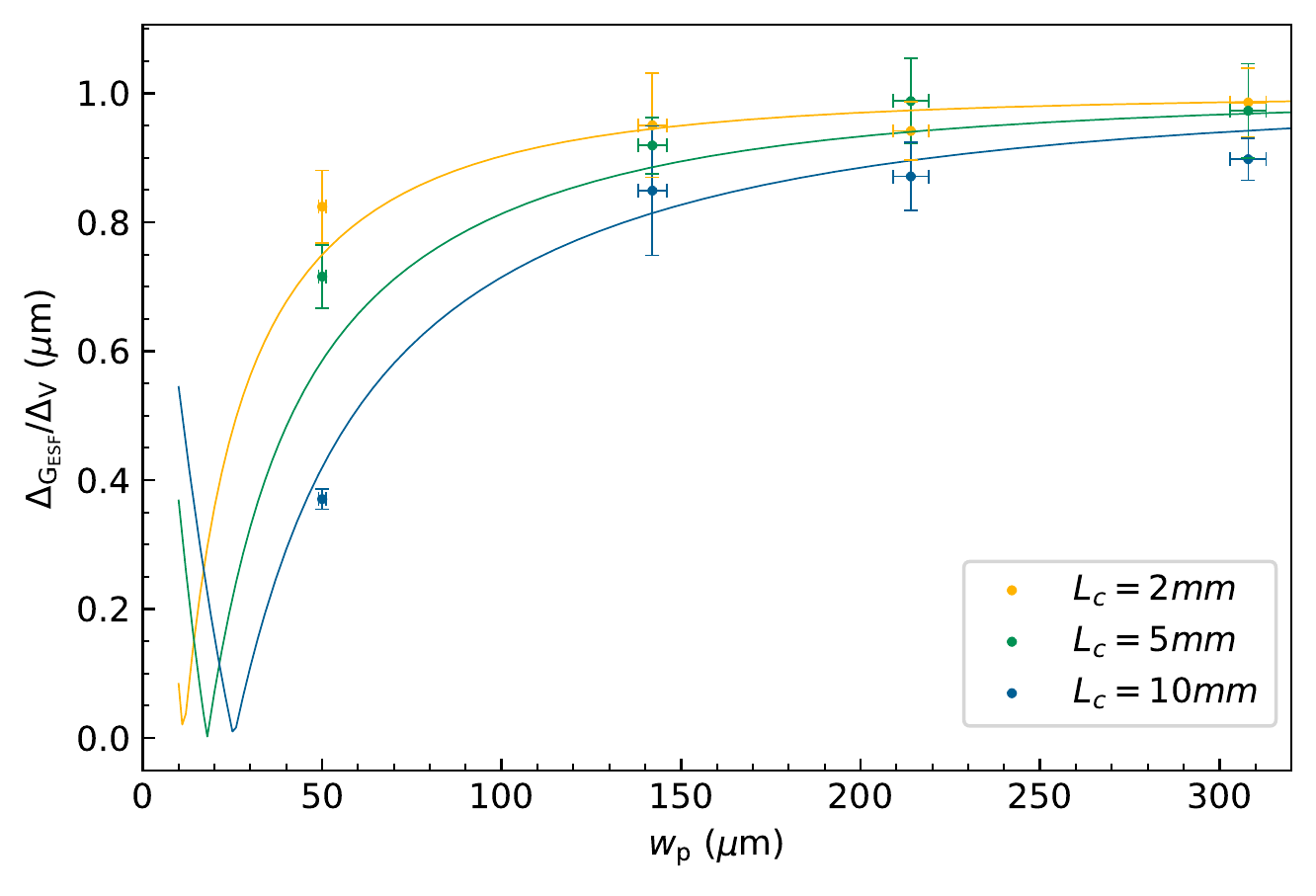}
	\caption{\textbf{Magnification independent quality parameter.} Ratio between the spread of the ESF derivative obtained from amplitude images ($\Delta_{\mathrm{G_{ESF}}}$) of a sharp edge and the corresponding spread obtained from visibility images ($\Delta_{\rmV}$). The solid lines show the theoretical predictions for each crystal length ($2\mm$ in yellow; $5\mm$ in green; $10\mm$ in blue) when varying the pump waist. Since the experimental points fit the theory prediction, we conclude that the underlying physics is well described by our theory model and that the experimental uncertainties are kept within acceptable limits. Because this ratio does not depend on the system magnification, it is a useful quantity to describe the system performance purely induced by the underlying quantum mechanical principles of the imaging scheme.}
 	\label{fig:ratio} 
\end{figure}

From the measured $\Delta_{\rmG,\rmc}$ and $\Delta_{\rmV,\rmc}$, the magnification independent parameter introduced in Eq.~\eqref{eq:ratio-M-independent} is calculated. We find that the percentage the experimental data deviates from the theory prediction is similar than the ratio between the extracted fit parameters $M_\rmd$ from Eq.~\eqref{eq:magnification-G} or Eq.~\eqref{eq:magnification-V}. Therefore, we can use this ratio indeed as a classifier to determine the quality of the experimental implementation to match the underlying theory assumptions used to model the system as described at the end of Sec.~\ref{sec:experimental setup}. 
Figure~\ref{fig:ratio} compares the experimentally measured and theoretically predicted ratios. The fact that they are in good agreement allows us to extract an estimator for the magnification of the detected interferometer arm for each of these measurements. Therefore, we have access to the magnification-adjusted spreads encoding the influence of the quantum correlations on the resolution. 

\begin{figure}
	\centering
		\includegraphics[width=0.8\linewidth]{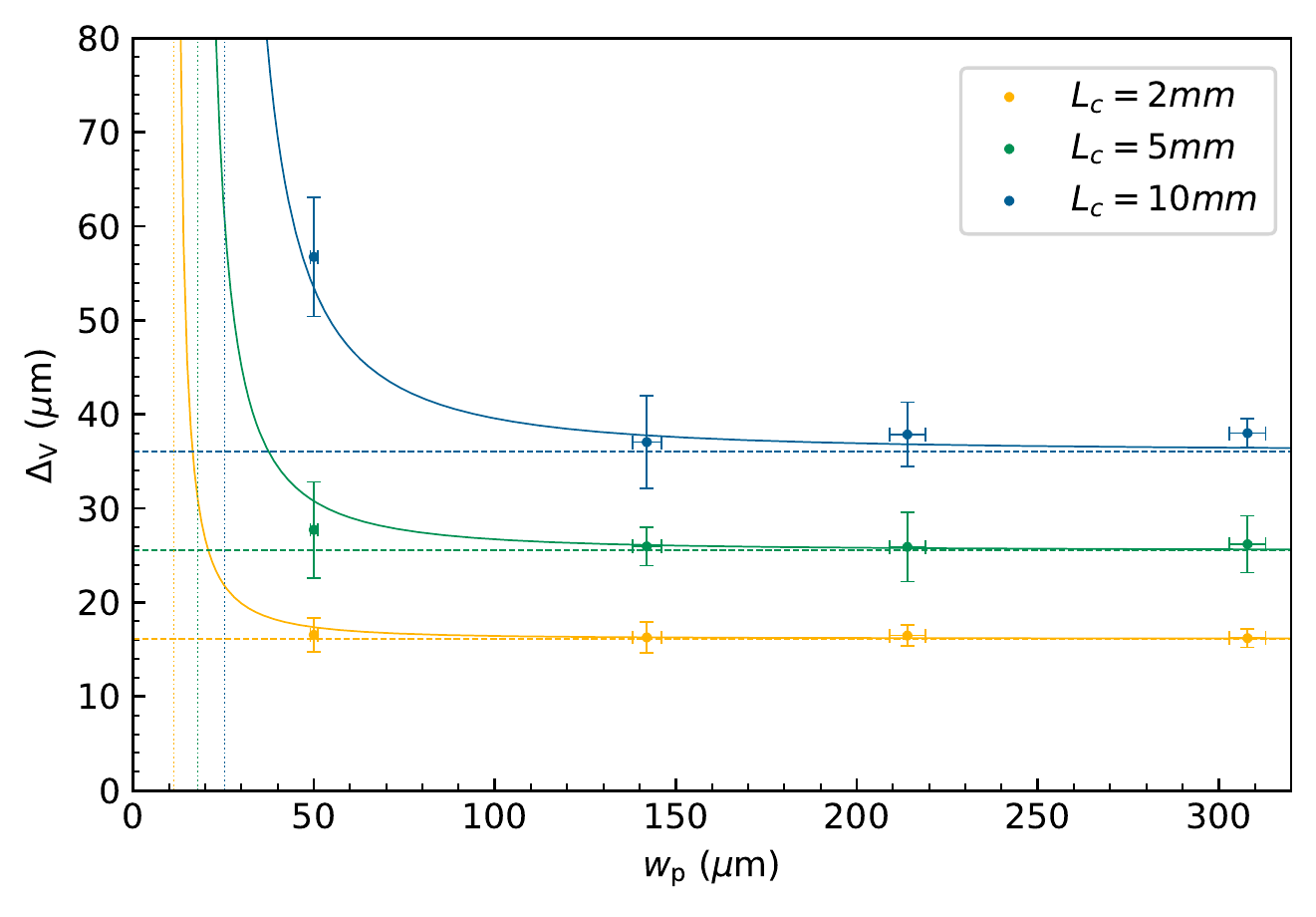}
	\caption{\textbf{Magnification adjusted visibility spreads for images taken through position correlations existing between SPDC photon pairs for a sharp edge.} Spread measured from visibility images (resolution) of a sharp edge for three crystals with different lengths ($2\mm$ in yellow; $5\mm$ in green; $10\mm$ in blue) when varying the pump waist. Experimental data is given in circles and compared to the theory prediction (solid lines of the same color). For comparison, we plot the limiting case of large pump waists as predicted in the literature \cite{Viswanathan2021_November} with dashed lines. For regimes where $\wP^2 \gg \frac{\lamd^2 L}{\lamd + \lamu}$ our extended theory model (see Sec.~\ref{sec:theory}) converges towards the simplified one, and resolution is mainly dependent on the crystal length. However, for smaller pump waists, the resolution worsens as the spatial correlations stored in the SPDC state worsen such that the biphoton state even becomes separable, i.e. the spatial correlations are lost for a specific parameter configuration (marked with dotted vertical lines for each crystal length).}
 	\label{fig:visPSF}
\end{figure}

\begin{figure}
	\centering
		\includegraphics[width=0.8\linewidth]{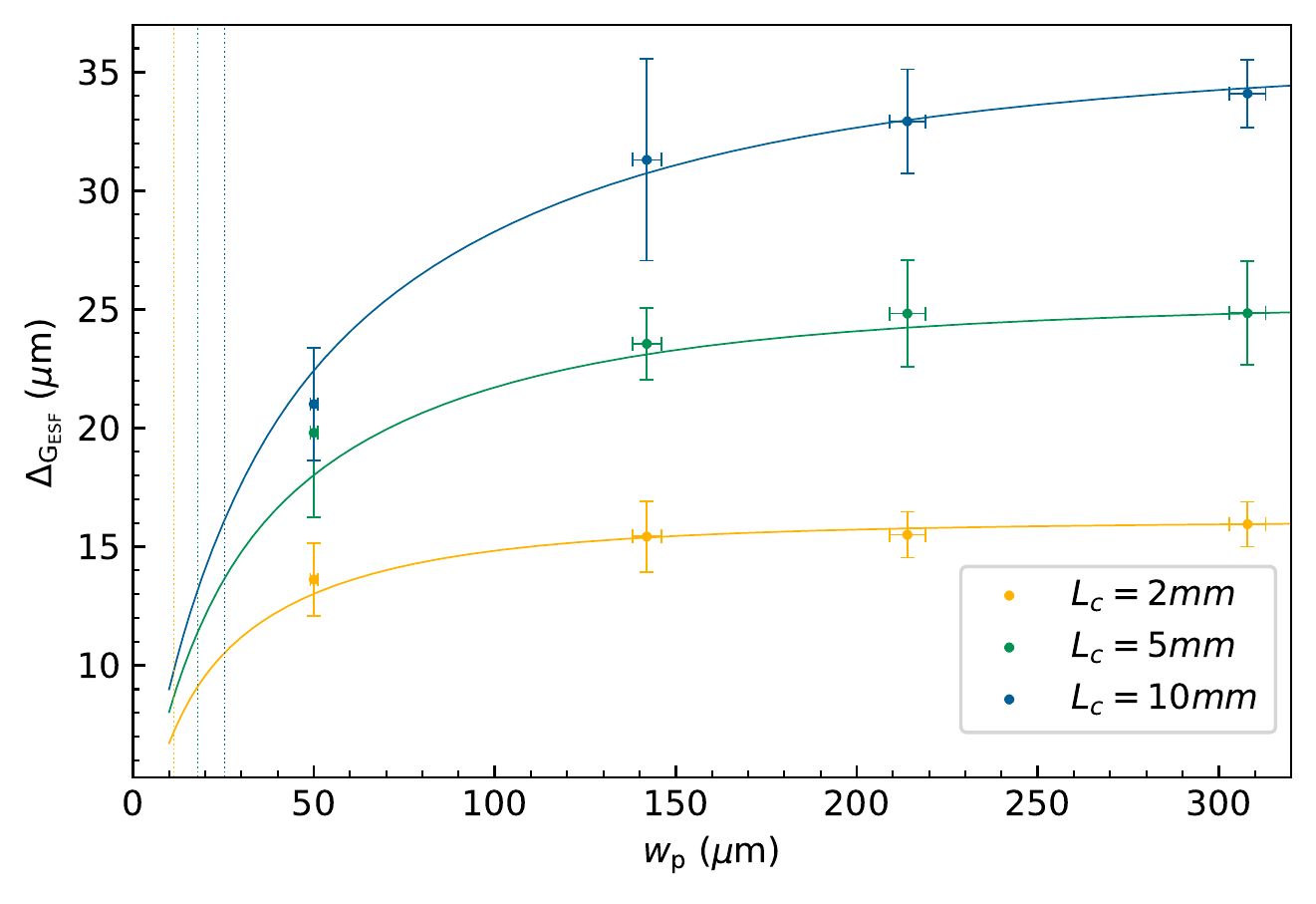}
	\caption{\textbf{Magnification adjusted spread $\Delta_{\rmG_{\ESF}}$ from amplitude images.} In case the pump waist gets smaller, the information on the object gets gradually erased between detected and undetected beams. At the point where the SPDC state becomes separable (marked with yellow dotted lines for a $2\mm$ crystal, in green for $5\mm$, and in blue for  $10\mm$), the amplitude images only carry information on the detected beam which does not contain any spatial information about the object. The spread measured is then related to the detected beam size. For larger pump waists ($\wP^2 \gg \frac{\lamu^2 L}{\lamd + \lamu}$), the derivative of the amplitude ESF can be considered as a good approximation of the LSF which can be related to the PSF. Moreover, the image and visibility functions are approaching the same limiting function in the large pump waist regime. Therefore, for this regime, the ESF derivative value converges towards the resolution value given in Fig.~\ref{fig:visPSF} determined by the visibility images.}
 	\label{fig:ampESF}
\end{figure}

The results from the evaluation of the magnification-adjusted spreads $\Delta_{\rmV}$ and $\Delta_{\rmG}$ as detailed in Sec.~\ref{sec:experimental setup} are given in Figures~\ref{fig:visPSF} and~\ref{fig:ampESF}, respectively. From the careful interpretation of these results, it is concluded that the physical meaning of what these two quantities encode is different. 
Visibility gives a measure on the indistinguishability of the beams and the correlation strength between the photons of an SPDC pair. These two quantities directly correlate to how good a point in the object is mapped onto the camera plane, i.e. they determine the image quality (resolution and contrast). This connection is also seen from the results in Fig.~\ref{fig:visPSF}, which show that the resolution of the system improves for short crystals, and stays constant when varying the pump waist  as predicted in the existing literature~\cite{Viswanathan2021_November, Vega2022}, if the pump waist is sufficiently large. To be more precise, we are able to identify this regime in the region of parameter space where $\wP^2 \gg \frac{\lamu^2 L}{\lamd + \lamu}$ and $\wP^2 \gg \frac{\lamd^2 L}{\lamd + \lamu}$ hold. However, an interesting behaviour arises for smaller pump waists. While decreasing the pump waist, the position correlations between the photon pairs deteriorate (which directly worsens image resolution as well) until they might be non-existing anymore, i.e. the SPDC biphoton state becomes separable. At that particular point, the resolution (or the visibility image PSF spread) diverges since the photons reaching the camera plane carry no spatial information on the object anymore. In order to explain the behaviour observed for smaller pump waists ($50\um$), it is necessary to extend the current existing models. This can be done either by following the lines of Ref.~\cite{Vega2022} using numerical techniques or by extending the existing analytical model as we will do in Sec.~\ref{sec:theory}.

For amplitude images, the interpretation of results presented in Fig.~\ref{fig:ampESF} has to be done more carefully. In Fig.~\ref{fig:ampESF}, we depict the spreads obtained from amplitude images depending on the pump waist. For large pump waists, one obtains similar results for visibility and amplitude images when analyzing image resolution. However, for smaller pump waists, when the position correlations start to worsen (see corresponding points in Fig.~\ref{fig:visPSF}), relating the amplitude image spread to resolution leads to misleading results. As the pump waist size approaches the value where the state becomes separable, the Gaussian contribution to the image function (see the exponential term in Eq.~\ref{eq:image_function}) induces the main $x_\rmc$ dependence compared to the contribution of the error function which approaches a constant value. At the point where the spatial correlations are lost, the image function carries no spatial information about the object. Therefore, the spreads obtained from amplitude images for small pump waists rather give a measure on the detected beam size than image resolution.

\section{Theory and discussion}
\label{sec:theory}

For the specific parameter constellations realized in the experiments, we observe that for large pump waists the resolution limits stay almost constant with varying pump waist. These results verify the theoretical predictions done in the literature so far that were operating in a regime where the influence of finite pump waists can almost be ignored \cite{Viswanathan2021_November, Vega2022}. However, the experimental results also demonstrate that discrepancies can arise if sufficiently small pump waists are realized for fixed crystal lengths $L$ and wavelengths $\lamd$ and $\lamu$. Even more interestingly, we observe that spreads obtained from visibility images or amplitude images have a different dependency on $\wP$. While both spreads approach the same limit for large pump waists, the definition of resolution in terms of the induced spreads via the imaging system becomes ambiguous for small pump waists as the image function spread $\Delta_\rmG$ decreases while the visibility spread $\Delta_\rmV$ increases. In order to address these subtle points, we are filling the gap of deriving an analytical model in the paraxial regime that takes the impact of the pump waist on the resolution limits into account. With that, we have a formalism for image formation with position correlations at our disposal such that we are able to analyze the resolution capabilities for a wide range of different source parameters based on the experimental setup sketched in Fig.~\ref{fig:1_NFsetup} as well as to identify the physical interpretation of $\Delta_\rmG$ and $\Delta_\rmV$.

One of the main ingredients for QIUL are the spatial correlations encoded in biphoton wave functions. Such correlated photon pairs are usually generated via SPDC and are the result of photons being born at approximately the same position \cite{Schneeloch}. In first order perturbation theory and for collinear phase matching the photon pair state reads ~\cite{Monken1998}
\begin{align}
 \ket{\psi} = \mathcal{N} \int \rmd \qd \int \rmd \qu \, P(\qd + \qu) \sinc \bigg(\frac{L \lambda_\mathrm{p} }{8\pi \lamd\lamu} \big(\lamd \qd - \lamu \qu \big)^2 \bigg) \ket{\qd}\ket{\qu}
\label{eq:SPDC-state}
\end{align}
where $\mathcal{N}$ is a normalization constant and $\qd$ ($\qu$) denotes the transverse wave vector of the detected (undetected) photon. Further, we have the wavelength of the pump photon $\lambda_\mathrm{p}$, the detected photon $\lamd$, and the undetected photon $\lamu$, the crystal length $L$ as well as the profile of a spatially coherent pump beam focused into the crystal. In our case, the latter is given by a Gaussian shape $P(\qd + \qu) = \exp\big\{ -\frac{w_\mathrm{p}^2}{4}(\qd + \qu)^2 \big\}$ with $w_\mathrm{p}$ being the pump waist. 

As we put the object at the image plane of the SPDC source, we exploit position correlations that are encoded in the joint probability density $\mathcal{P}(\rhobf_\mathrm{d},\rhobf_\mathrm{u})$.  
In order to analyze the properties of our QIUL setup, we use the image function $G(\rhoc)$ as well as the visibility $V(\rhoc)$, see Eq.~\eqref{eq:imageFunction-Def} and Eq.~\eqref{eq:visibility-Def}, respectively. Following Ref.~\cite{Viswanathan2021_November}, the image function can be computed in our specific case via
\begin{align}
    G(\rhoc) \sim \int \rmd\rhoo \, \mathcal{P}\left(\frac{\rhoc}{M_\rmd},\frac{\rhoo}{M_\rmu}\right) |T(\rhoo)|.
\label{eq:image_function}
\end{align}
Analogously, we have for the visibility
\begin{align}
    V(\rhoc) \sim \frac{\int \rmd\rhoo \, \mathcal{P}\left(\frac{\rhoc}{M_\rmd},\frac{\rhoo}{M_\rmu}\right) |T(\rhoo)|}{\int \rmd\rhoo \, \mathcal{P}\left(\frac{\rhoc}{M_\rmd},\frac{\rhoo}{M_\rmu}\right)}.
\label{eq:visibility}
\end{align}

The impact of an object is encoded in the transmission coefficient $T(\rhoo)$. Simple models for an object are given by a Dirac delta function, $T\sim \delta(\rhoo)$, modeling a point or a Heaviside function, $T = \Theta(x_\rmo)$, modeling the impact of an edge being orthogonal to the $x_{\rmo}$ direction in the object plane. 
We will denote the image function evaluated for the respective objects as $G_{\PSF}$ for a point and $G_{\ESF}$ for an edge. Similarly, we introduce the notation $V_{\PSF}$ (visibility PSF) and $V_{\ESF}$ (visibility ESF). 

Due to the intricate momentum dependency of the SPDC state~\eqref{eq:SPDC-state}, it is a nontrivial task to find a closed form expression for the joint probability density $\mathcal{P}(\rhobf_\rmd,\rhobf_\rmu)$ and thus for the image function or visibility. To obtain a qualitative understanding, we approximate the $\sinc$ by a Gaussian structure, $\sinc(x^2) \to \E^{-x^2}$, following the standard strategy usually done in the literature \cite{Viswanathan2021_November}. 
For this particular approximation, one obtains
\begin{align}
 \mathcal{P}(\rhobf_\rmd,\rhobf_\rmu) = 
 \frac{8}{\pi\wP^2 L (\lamd+\lamu)} \exp\left\{-\frac{2(\lamu \rhobf_\rmd + \lamd \rhobf_\rmu)^2}{\wP^2(\lamd+\lamu)^2} - \frac{4\pi (\rhobf_\rmd - \rhobf_\rmu)^2}{L(\lamd+\lamu)}\right\}
 \label{eq:JointProbPos}
\end{align}
for the joint probability density. So far, the resolution limit for the undetected photon scheme under investigation was analyzed in the limit where the first term in the exponential is merely slowly varying compared to the second term in the sum. Formally, this is equivalent with a plane-wave limit where $\wP \to \infty$. This is motivated by the fact that typical parameters realized in an experiment allow to neglect the contributions from a finite pump waist. Indeed, our experimental data clearly show that this is a well justified approximation over a large parameter range of the pump waist for fixed $\lamd$, $\lamu$, $L$. Nonetheless, we also demonstrated that we are able to probe regimes where the pump waist influences the imaging system. Therefore, we extend the existing analyses by including the impact of finite pump waists. 

As a first example to describe the resolving power of the optical system sketched in Fig.~\ref{fig:1_NFsetup} in a qualitative fashion, we analyze the PSF for the Gaussian approximation of the $\sinc$ function and obtain
\begin{align}
 G_{\PSF}(\rho_\rmc) &= \exp\left\{ -\bigg[ \frac{2\lamu^2}{\wP^{2}(\lamd + \lamu)^2} + \frac{4\pi}{L (\lamd + \lamu)} \bigg] \frac{\rho_\rmc^2}{M_\rmd^2} \right\}, \intertext{as well as}
 V_{\PSF} (\rho_\rmc) &= \exp\left\{ - \frac{2\big[2\pi\wP^2(\lamd+\lamu) - \lamd\lamu L \big]^2}{2\pi\wP^4(\lamd+\lamu)^3 L  + \wP^2 \lamd^2 (\lamd+\lamu)^2 L^2} \frac{\rho_\rmc^2}{M_\rmd^2} \right\}
 \label{eq:VisPSF}
\end{align}
for the image function PSF and visibility PSF, respectively. Here, we have used the fact that the PSFs obey a radial symmetry, thus, depending only on $\rho_\rmc = |\rhoc|$. Furthermore, we normalized the maximum to one. 

Usually, the quality of a QIUL system is quantified by the spreads of the image function or visibility. As aforementioned, we are using a $1/\E$-width for the PSFs, $G_{\PSF}(\Delta_{\rmG_{\PSF},\rmc}) = 1/\E$ and $V_{\PSF}(\Delta_{\rmV,\rmc}) = 1/\E$. Note, that we have introduced a subscript $\PSF$ for the spread of the image function to indicate that this is a spread obtained from a PSF but dropped it for the visibility spread. The reason for this will become clear once we discuss the spreads of the ESFs. Eventually, we obtain the magnification adjusted PSF spreads by dividing the PSF spreads at the camera by the magnification of the detected arm, $\Delta_{\rmG_{\PSF}} = \Delta_{\rmG_{\PSF},\rmc}/M_\rmd$ and $\Delta_{\rmV} = \Delta_{\rmV,\rmc}/M_\rmd$, which read
\begin{align}
 \Delta_{\rmG_{\PSF}} &= \sqrt{ \frac{L(\lamd + \lamu)}{4\pi}}\; \sqrt{ \frac{1}{1 + \frac{\lamu^2 L}{2\pi \wP^2 (\lamd + \lamu)}}},
\label{eq:spread-G} \\
 \Delta_{\rmV} &= 
  \sqrt{ \frac{L(\lamd + \lamu)}{4\pi}}\; 
  \frac{2\pi\wP^2 (\lamd + \lamu) \sqrt{1 + \frac{\lamd^2 L}{2\pi\wP^2 (\lamd + \lamu)}}}{2\pi \wP^2 (\lamd + \lamu)-\lamd\lamu L}.
\label{eq:spread-V}
\end{align}
The first conclusion that we can draw from Eq.~\eqref{eq:spread-G} and \eqref{eq:spread-V} is that the spreads obtained from the image function as well as from the visibility coincide in the $\wP \to \infty$ limit. In particular, $\Delta_{\rmG_{\PSF}}$ coincides with the result in Ref.~\cite{Viswanathan2021_November} in this limit where contributions of the pump waist on the resolution were neglected. For the image function spread, this is a good approximation as long as the inequality $\frac{\lamu^2 L}{2\pi \wP^2 (\lamd + \lamu)} \ll 1$ is fulfilled. Interestingly, the two spreads either obtained from the image function or from visibility show different dependencies on $\wP$. For instance, $\Delta_{\rmV}$ is well approximated by the pump waist independent limit $\sqrt{ \frac{L(\lamd + \lamu)}{4\pi}}$ for $\frac{\lamd^2 L}{2\pi \wP^2 (\lamd + \lamu)} \ll 1$ and $\frac{\lamd\lamu L}{2\pi \wP^2 (\lamd + \lamu)} \ll 1$. Apart from the limiting case, there are a couple of important differences stored in both quantities that become manifest for finite pump waists. 
In case the pump waist is decreasing (for fixed other parameters), the values for $\Delta_{\rmV}$ increase until they reach a singularity at $w_{\mathrm{p,sing}}^2 = \frac{\lamd\lamu L}{2\pi (\lamd + \lamu)}$, cf. Fig.~\ref{fig:visPSF}. By contrast, $\Delta_{\rmG_{\PSF}}$ is decreasing. Naively, one could conclude that the resolution improves with smaller pump waists by investigating the image function spread. However, the spread extracted from the visibility gets broader for smaller pump waists until it hits a singularity within our approximation. While this seems to be a contradiction at first sight, it is important to notice that both quantities store different information of the presented imaging scheme. 

The fact that $\Delta_{\rmV}$ diverges is not surprising if we carefully study the properties of the SPDC biphoton state enabling imaging with undetected photons. If the condition $\wP = w_{\mathrm{p,sing}}$ is fulfilled, the biphoton state becomes separable within the Gaussian approximation of the $\sinc$ function. Even more important, all spatial correlations between the detected and undetected photons are lost. This becomes transparent if one studies the joint probability distribution given in Eq.~\eqref{eq:JointProbPos} which factorizes $\mathcal{P}(\rhobf_\rmd,\rhobf_\rmu) = \mathcal{P}_\rmd(\rhobf_\rmd) \mathcal{P}_\rmu(\rhobf_\rmu)$. As both photons of a pair are uncorrelated, the spatial information cannot be transmitted from the object to the camera. Therefore, the visibility becomes constant, cf. Eq.~\eqref{eq:VisPSF} for $\wP = w_{\mathrm{p,sing}}$, and $\Delta_{\rmV}$ diverges. Thus, $\Delta_{\rmV}$ can be interpreted as a measure of the correlation strengths of the biphoton state. For large pump waists, there exist a high degree of spatial correlations. Lowering the pump waist, the correlations get worse until they vanish at the singularity. Technically speaking, the correlation strengths begin to increase again for $\wP < w_{\mathrm{p,sing}}$. However, we might also approach a regime there were our assumptions, e.g. paraxial approximation, break down. Furthermore, it is important to note that the singularity might be an artefact of the Gaussian approximation. By taking the actual $\sinc$ phase matching condition into account, we assume that the singular structure might get softened, depending as to whether a parameter combination exists such that the actual SPDC state given in Eq.~\eqref{eq:SPDC-state} gets separable.

The role of the image function spread $\Delta_{\rmG_{\PSF}}$ is different. The image function encodes intensities at each camera pixel position. Even though the biphoton state might become separable for a specific parameter constellation, there is always the detected beam impinging onto the camera. In our currently analyzed case, it will have a Gaussian shape with a spread determined by $M_\rmd\sqrt{\frac{L\lamd}{4\pi}}$ within our approximations. However, this spread is not a valid measure to quantify the spatial resolution capabilities of the undetected photon scheme. The detected photons in this case do not contain any spatial information of the object at all as the image function does not properly reflect the correlation strengths in an adequate manner. Therefore for wide-field imaging, we are using the spreads extracted from visibility information to quantify resolution limits. The image function spread rather provides information about the detected beam size. Only in the limit $\wP \to \infty$, image function and visibility store the same information as in this case the SPDC state becomes perfectly correlated.

Eventually, the (visibility) spread observed in the detection plane divided by the total system magnification $\frac{M_\rmd}{M_\rmu}$ can be related to the minimum resolvable distance of an object via \cite{Viswanathan2021_November}
\begin{equation}
    d_\mathrm{{min, obj}_{(NF)}} \approx
    0.7\sqrt{2\pi}\frac{M_\rmu}{M_\rmd}\Delta_{\rmV,\rmc}
    = 0.7 M_\rmu \sqrt{ \frac{L(\lamd + \lamu)}{2}}\; 
  \frac{2\pi\wP^2 (\lamd + \lamu) \sqrt{1 + \frac{\lamd^2 L}{2\pi\wP^2 (\lamd + \lamu)}}}{2\pi \wP^2 (\lamd + \lamu)-\lamd\lamu L}.
    \label{eq:near-field_resolution}
\end{equation} 
One can deduce from these analytical solutions that the dominating effects for the resolution are given by the crystal length $L$ and the magnification of the undetected arm $M_\rmu$ for a wide range of pump waists. Although, the terminology of a magnification is used for the quantity $M_\rmu$, we would like to emphasize that it does not play the role of an usual magnification as in a classical imaging scheme. The system of lenses inducing $M_\rmu$ rather decreasing ($M_\rmu <1$) or increasing ($M_\rmu >1$) the illumination spot of the undetected photon beam but do not magnify any properties of the object. Therefore, the parameter $M_\rmu$ influences the actual resolution while the parameter $M_\rmd$ indeed plays the role of a magnification as it is magnifying the spatial information in the detected light beam transmitted via the correlations stored in the joint probability density from the undetected photons interacting with the object. 

In general, the resolution improves with shorter crystal lengths being consistent with the fact that this implies stronger position correlations as can be seen in Eq.~\eqref{eq:JointProbPos}. However, this improvement of the resolution is limited by a threshold determined by the longer wave length of the SPDC generated photon pairs. This effect is not present in the current model as we perform momentum integrations over the entire momentum space of detected and undetected photons. In practice, the available propagating modes are constraint, thus, modifying the integration boundaries. As long as $L\gg \lamd+\lamu$ these effects can be neglected but in case $L \approx \lamd+\lamu$, the localisation of the PSF for decreasing $L$ saturates, resulting in a crystal-length independent spread for thin crystals with $L < \lamd+\lamu$. This effect was analyzed and corroborated by detailed numerical studies~\cite{Vega2022}.

While we have theoretically analyzed the resolution properties based on position correlations for the simple case of a point as an object, we now extend the analysis to the situation of a sharp edge to be able to compare to the experimental results presented in Sec.~\ref{sec:results}. Therefore, we evaluate Eqs.~\eqref{eq:image_function} and \eqref{eq:visibility} for $T=\Theta(x_\rmo-\tilde{x}_\rmo)$ leading to $G_\ESF$ and $V_\ESF$ given in Eqs.~\eqref{eq:magnification-G} and \eqref{eq:magnification-V}, respectively. On a qualitative level, we can draw the same conclusions from the ESFs as we did for the PSFs. For the visibility this extends even to a quantitative level. This can directly be inferred from the fact that in our case the derivative of the visibility ESF with respect to the position in the camera plane is mathematically equivalent to our result of the visibility PSF $\partial_{x_\rmc} V_\ESF = V_\PSF$. Therefore, we extract the same spreads according to our resolution criteria specified in Sec.~\ref{sec:experimental setup}. For the image function, the situation is slightly different as $\partial_{x_\rmc} G_\ESF \neq G_\PSF$ due to the fact that the joint probability density is not only a function of the coordinate differences. Due to the nontrivial $x_\rmc$ dependency of $G_\ESF$ there is no closed form expression for the image function ESF spread $\Delta_{\rmG_{\ESF}}$. Nevertheless, we can extract this information numerically which is depicted as solid lines in Fig.~\ref{fig:ampESF}.

\section{Conclusion}
We have experimentally evaluated for the first time the effect of crystal length and pump waist on image resolution in a quantum imaging system exploiting position correlations of down-converted photons. The results obtained confirm the theory predictions published so far in the regime where nontrivial effects from the pump waist can be neglected. Nevertheless, the experimental results clearly demonstrated deviations from the existing analytical predictions for decreasing pump waists if the other parameters where kept fixed. We therefore derived an analytical model to predict the resolution values as well as the impact of the correlation strength over a larger parameter range. 

Moreover, we analyzed the physical meaning encoded in visibility and amplitude image information.
We found that visibility is the property containing the resolution information while amplitude images give rather information on the detected beam size. For the regime of small pump waists, spatial resolution worsens as the correlations between the photons of the biphoton state deteriorate. That results in the fact that wide-field imaging being not a suitable approach in case the spatial correlations between the photons are lost. However, QIUL would still be possible in this regime by using a scanning approach. In this case, the image resolution would not depend on quantum properties but rather depend on the bit depth of the camera, the scanning step size, and the undetected beam size.

To summarize, an improvement on image resolution can be mainly achieved by using shorter crystal lengths and by decreasing the magnification at the undetected path ($M_\rmu$) as previously stated \cite{Viswanathan2021_November}. At the same time, it is important to stress out that $M_\rmu$ does not act as a classical magnification since it does not influence the detected dimensions on the camera, but just modifies the spot size that probes the object. In addition, our results show that resolution gets worse as soon as the pump waist leaves the regime where $\wP^2 \gg \frac{\lamd\lamu L}{\lamd+\lamu}$ and $\wP^2 \gg \frac{\lamd^2 L}{\lamd+\lamu}$ are fulfilled.

Our work provides new insights into the intricate relations between all source parameters and properties, providing us with a new tool to optimize image resolution for different imaging applications.


\section{Acknowledgements}
This work was supported as a Fraunhofer LIGHTHOUSE PROJECT (QUILT). 
Furthermore, funding from the German Federal Ministry of Education and Research (BMBF) within the funding program Quantentechnologien - von den Grundlagen zum Markt with contract number 13N16496 as well as the funding programme FKZ 13N14877 are acknowledged. We also acknowledge support from the Thuringian Ministry for Economy, Science, and Digital Society and the European Social Funds (2021 FGI 0043); European Union's Horizon 2020 research and innovation programme (Grant Agreement No. 899580 and No. 899824); and the Cluster of Excellence "Balance of the Microverse" (EXC 2051 project 390713860).

\section{Appendix}
\label{Appendix}

\subsection{Imaging configuration}
\label{app:imaging_configuration}
\begin{figure}
	\centering
		\includegraphics[width=0.9\linewidth]{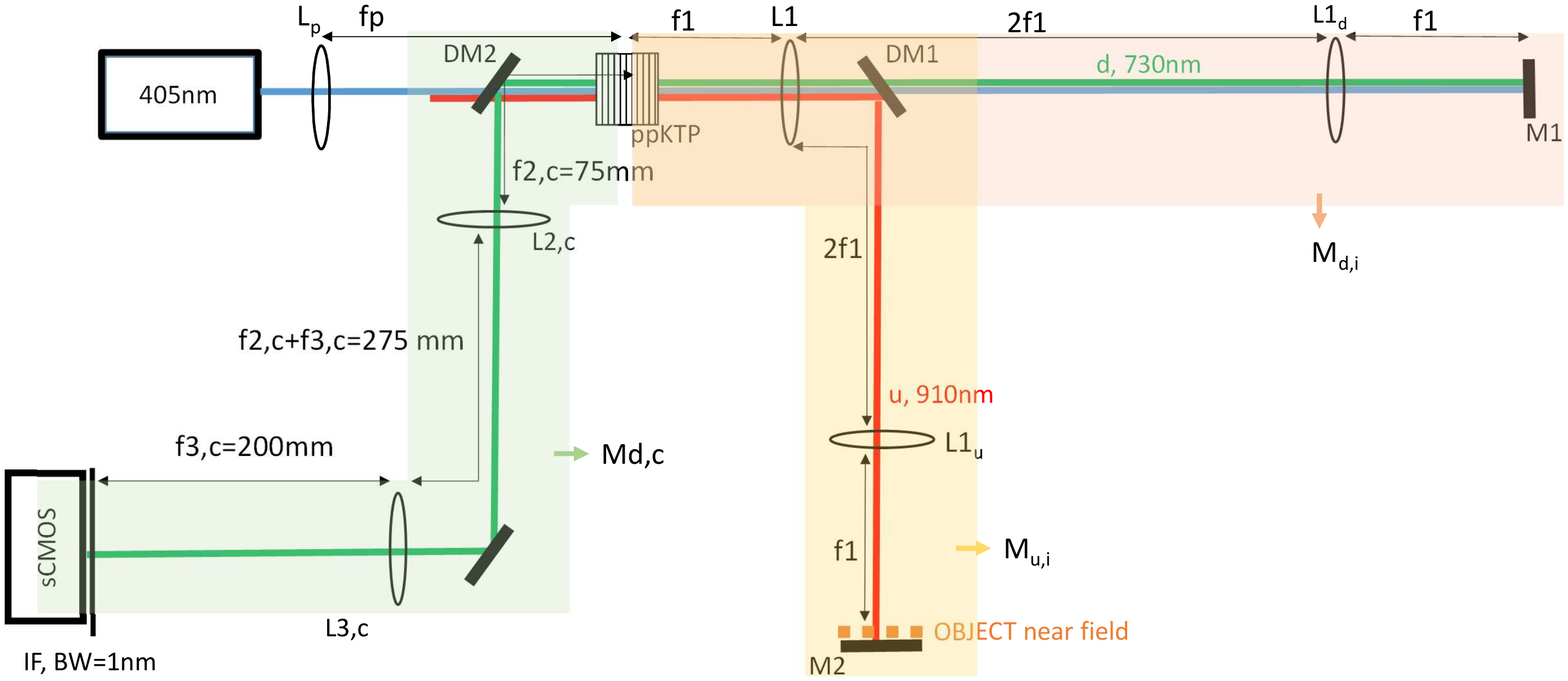}
    \caption{\textbf{Near-field configuration setup.} Sketch of the setup used for this work with detailed information on the lenses used and distance between them in order to exploit position correlations between the down-converted photons. Lenses are noted with $\mathrm{L}$, dichroic mirrors with $\mathrm{DM}$, $\mathrm{M1}$ and $\mathrm{M2}$ are the mirrors of the detected and undetected interferometer arms, respectively. Detected and undetected paths magnification value inside the interferometer are noted as $\mathrm{M}_{\rmd,\mathrm{i}}$ and $\mathrm{M}_{\rmu,\mathrm{i}}$, while $\mathrm{M}_{\rmd,\rmc}$ refers to the magnification seen by the detected beam between the ppKTP crystal and the camera. Distances between the lenses are shown with arrows and specified in terms of the focal length of the corresponding lenses. Before the camera, we have an interference filter (IF) with $1.5\nm$ bandwidth (BW).}
	\label{fig:1_NFsetup_simple}
\end{figure}

Figure \ref{fig:1_NFsetup_simple} gives detailed information on the imaging configuration used during the measurements presented. 

\subsection{Magnification measurement}
\label{app:MagMeas}

\begin{figure}
	\centering
		\includegraphics[width=0.8\linewidth]{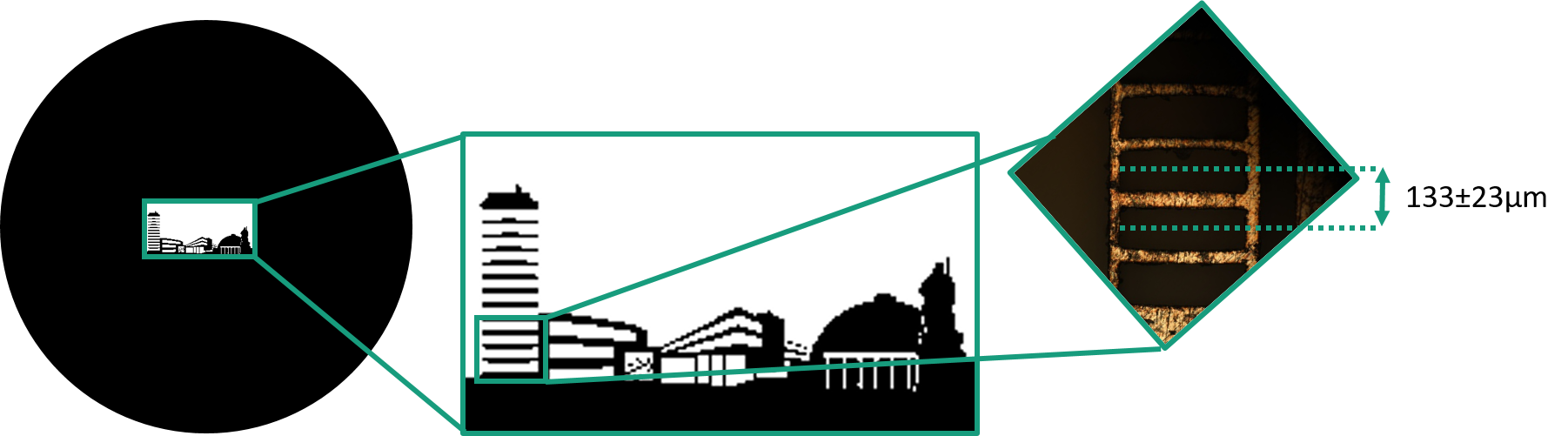}
	\caption{\textbf{In-house made object used for the magnification measurements.} The object is laser cut on a thin metal sheet. Therefore its transmission is either 0 or 1 for the probing light. It represents the Jena skyline, and the tower windows are used as slits for the magnification measurements. The distance between the center of the windows (slit distance) is measured to be $133\pm 23\um$. Where the uncertainty of the distance corresponds to the sum of the measuring device uncertainty and the manufacturing precision.}
	\label{fig:3_skyline}
\end{figure}

\begin{figure}
	\centering
		\includegraphics[width=0.8\linewidth]{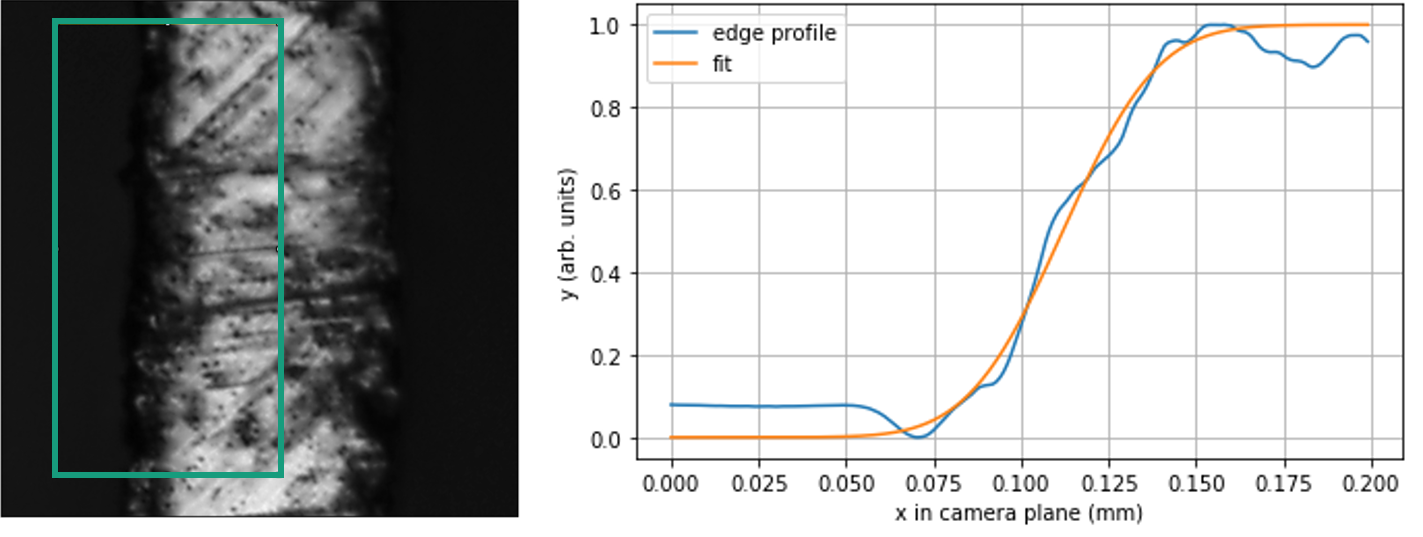}
	\caption{\textbf{Object manufacturing precision measurement.} The image of a window edge was analyzed by plotting the averaged intensity edge profile along the x-axis of the area marked in green. That profile is then fitted with an error function and its spread evaluated at 1/e is taken as the manufacturing precision of one window edge. Notice that this value must be doubled to account for the two edges.}
	\label{fig:4_object_error}
\end{figure}

An in-house made object (Fig. \ref{fig:3_skyline}) was used to perform the magnification measurements by calculating the ratio between the object and its image dimensions. Due to the field of view (FOV) at the object plane, only the "windows of the tower" acting as parallel slits with a fixed distance are considered as our object. The object dimensions were measured with a Zygo- New View 7300 optical profiler. The roughness of the frame between the windows is the main error source of the object dimension measurements, and therefore, it was also measured. The manufacturing precision of the object was analysed from a picture taken under a $20\times$ magnification objective with an Olympus DP71 sensor which is coupled to an Olympus BX51TRF microscope with a U-TV0.63XC adapter. The unsharpness of one edge of the window frame is obtained by fitting an error function to its intensity profile (Fig. \ref{fig:4_object_error}). From these measurements, we obtain that the distance between two window centers is $133\um \pm 23\um$.

\begin{figure}
	\centering
		\includegraphics[width=0.8\linewidth]{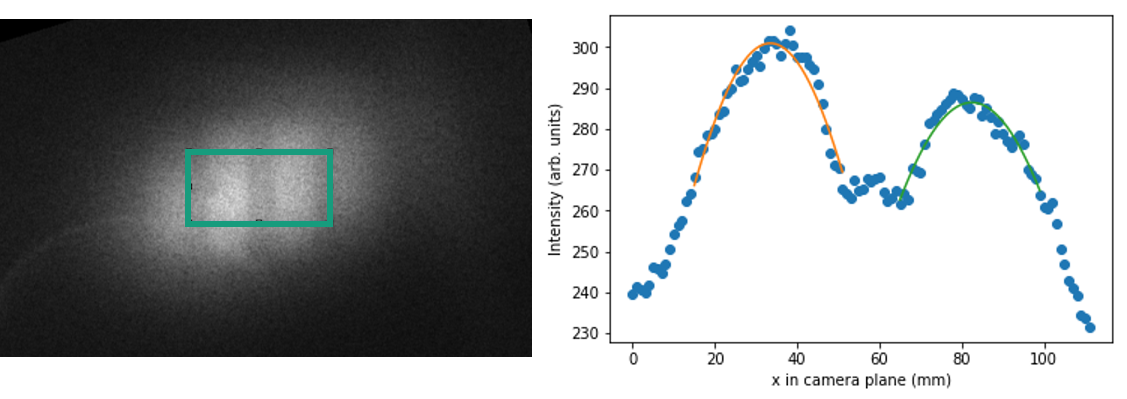}
	\caption{\textbf{Object image dimensions analysis.} Image of the tower windows taken with our setup but illuminating the object with the detected wavelength to obtain a classical image. The windows are treated as slits and their intensity profile are fitted by Gaussian functions. The distance between two slits (windows) is then taken as the distance between the maxima of the Gaussian fits. This example is taken with a $5\mm$ long crystal and a pump waist of $214\um$.}
	\label{fig:5_image_error}
\end{figure}

To measure the distance between two windows on the image obtained with the QIUL system, the windows of the tower are treated as slits. For these measurements, the wavelength that illuminates the object is the same as the detected, and the possibility to create light from the second pass of the pump through the crystal is avoided by blocking that path. The intensity profile from each slit (window) was fitted with a Gaussian function, and the distance between slits (windows) is then the distance between the Gaussian peaks given in pixel units (Fig. \ref{fig:5_image_error}). This distance is then converted to $\um$ from the camera sCMOS camera pixel size of $6.5\um$. By taking this measurement right after the resolution measurements for each crystal length and pump waist combination, and comparing it to the object real dimensions, we calculated the experimental magnification values for each configuration.

\section{References}
\bibliography{references}

\end{document}